\documentclass{article}
\usepackage{graphicx} 
\usepackage{geometry}
\usepackage{setspace}
\usepackage{float}
\usepackage{multirow}
\usepackage{multicol}
\usepackage{lipsum}
\usepackage{mwe}
\usepackage{sidecap}
\begin{Huge}
\title{Developing a Novel Holistic, Personalized Dementia Risk Prediction Model via Integration of Machine Learning and Network Systems Biology Approaches}
\end{Huge}
\author{Srilekha Mamidala}
\date{}

\begin{document}

\maketitle

\newpage
\begin{spacing}{1.5}
\section{Abstract}

The prevalence of dementia has increased over time as global life expectancy improves and populations age. An individual’s risk of developing dementia is influenced by various genetic, lifestyle, and environmental factors, among others. Predicting dementia risk may enable individuals to employ mitigation strategies or lifestyle changes to delay dementia onset. Current computational approaches to dementia prediction only return risk upon narrow categories of variables and do not account for interactions between different risk variables. The proposed framework utilizes a novel holistic approach to dementia risk prediction and is the first to incorporate various sources of tabular environmental pollution and lifestyle factor data with network systems biology-based genetic data. LightGBM gradient boosting was employed to ensure validity of included factors. This approach successfully models interactions between variables through an original weighted integration method coined \textit{Sysable}. Multiple machine learning models trained the algorithm to reduce reliance on a single model. The developed approach surpassed all existing dementia risk prediction approaches, with a sensitivity of 85\%, specificity of 99\%, geometric accuracy of 92\%, and AUROC of 91.7\%. A transfer learning model was implemented as well. De-biasing algorithms were run on the model via the AI Fairness 360 Library. Effects of demographic disparities on dementia prevalence were analyzed to potentially highlight areas in need and promote equitable and accessible care. The resulting model was additionally integrated into a user-friendly app providing holistic predictions and personalized risk mitigation strategies. The developed model successfully employs holistic computational dementia risk prediction for clinical use.

\newpage
\section{Introduction}

By 2030, 1 in 6 people are predicted to be 60 years or older [48]. The population of senior citizens has dramatically increased through a rise in human life expectancy and historical trends that have been expected to continue until at least 2060 with a growing middle-aged population [48, 52]. However, with this trend comes a heightened risk for cognitive impairment among these populations. Severe cognitive impairment (SCI) such as dementia has multiple consequences, including the mental, emotional, and financial tolls such conditions have on patients. The prevalence of SCI also increases rapidly with age, with approximately 2\% of senior citizens aged 65-69 and 30-50\% of people over the age of 90 reporting SCI, and the number of current middle-aged individuals expected to report an even higher prevalence of SCI in coming years [51]. 

This paper focuses on predicting risk of dementia, a form of SCI. Dementia is the general impairment of cognitive functions such as speaking or thinking [38]. Dementia is not a normal part of the aging process and differs from the mild memory loss commonly associated with aging as a person with dementia may not be able to engage in regular occupational activities and may require assisted care since it interferes with daily life [39]. Dementia is generally known as the damage of neuron connections in the brain, but the underlying causes are unknown [38].  

The most common form of dementia is Alzheimer’s Disease. A neurodegenerative disorder that slowly destroys a patient’s memory and thinking skills, Alzheimer's Disease has affected more than 6 million people aged 65 years and older in the United States alone [37]. Alzheimer’s Disease often reveals itself in the form of amyloid plaques, abnormally high buildups of the protein beta-amyloid that form plaques when they collect near neurons and interfere with regular cellular processes such as the loss of neuron connections in the brain. To refer to the umbrella of dementia-related disorders, the term Alzheimer's disease and related dementias (ADRD) is used [37]. Predicting individualized risk for ADRD through the following factors can provide insight for middle-aged and senior citizens, and can motivate them to see a medical professional for further guidance and a possible diagnosis. 

There are several factors affecting the risk of developing SCI. Environmental factors such as air pollution have been shown to impact cognitive impairment. Poor quality air containing compounds such as nitrogen oxides and PM 2.5 particulate matter has shown a positive correlation with dementia cases [57]. Studies have shown that prolonged exposure to air pollution resulted in higher beta amyloid protein levels in the blood, a protein commonly associated with cognitive disorder [11, 47, 57]. 

Genetic variants related to neurotransmitters and neurological function can affect executive functioning and memory. 1 in 4 people aged 55 years or older has a close relative with dementia, revealing that genetics may play a role in determining how much an individual is at risk for the disease [10]. Analyzing gene variants, or the different versions of a gene that vary from person to person, can factor into cognitive disorder risk. For example, three of the known gene variants that affect Alzheimer’s Disease are APOE2, APOE3 and APOE4, the last being the greatest contributor to Alzheimer’s Disease risk [3, 10, 37].

Lifestyle factors such as diet, social engagement, smoking, and excessive alcohol consumption also affect the risk of developing a cognitive disorder [4, 41]. Regular exercise and the maintaining of a healthy, balanced diet promotes cognitive health, as well as overall health [41]. 

This project utilized a systems biology and machine learning based approach to predict risk for dementia. Systems biology is the modeling of complex interactions via networks within biological systems to understand the system as a whole. It was chosen to model the genetic interactions to determine risk because the complex, multi-faceted nature of these interactions can most accurately be modeled through a systems biology network [22]. Machine learning is the ability of a computer-based algorithm to learn patterns from successive iterations of data training, similar to that of a human. This approach was chosen to optimize the accuracy of a risk prediction algorithm that relies on large sets of data that needs to be organized and learned. It was hypothesized that employing a systems biology and machine learning based approach would successfully predict risk for cognitive disorders because of its ability to factor genetic, environmental, and lifestyle influences to form a tailorable, connective network that is able to successively learn to predict risk more accurately. 

The use of computational approaches for disease risk prediction has greatly increased in recent years as machine learning technologies become more widely accessible. The majority of disease risk prediction models aim to predict risk for heart and cardiovascular diseases and their related counterparts. Dementia risk prediction models are, in comparison, less prevalent. Existing dementia risk prediction models have employed machine learning to explore dementia risk prediction through various factors, the majority concentrating on predicting risk from genetic data [14, 16, 22] through large, population based longitudinal studies [16, 43]. Furthermore, the machine learning approach of Support Vector Machines (SVM) has been predominantly used among existing papers [43, 51]. Tang et. al separated dementia risk models into four main categories: demographic-only models, health variable models, cognitive imaging/testing models, genetic models, and multivariable models, illustrating a wide variety of available model prevalence. Although existing models successfully train for a single category of variables, most are often unable to account for interactions between different risk variables because the sole use of machine learning methods does not permit this [19], and no other model combines environmental pollution data with genetic and lifestyle factor data. Multiple external literature reviews studying computational dementia prediction models have noted that the range of models drawing conclusions from only narrow categories of data have created challenges when comparing models as there are no models encompassing all types of factors [43, 51]. 

The innovation in the proposed model lies in the fact that it is the first computational method to integrate systems biology and machine learning methods to analyze the interactions between factors to produce a singular output of a personalized risk assessment. This algorithm is also the first outputting risk based on genetic, environmental, and lifestyle variables to create a unique holistic risk assessment tool. The algorithm was also developed into a user-friendly app to enable middle-aged and elderly persons to determine individual risk for dementia, with the app providing personalized risk mitigation strategies. Calculating dementia risk is not only pertinent for elderly populations: provided an individual remains in the same conditions and habits over longer periods of time, the proposed algorithm can predict potential dementia onset for up to 10 years beforehand. Therefore, it is imperative that middle-aged persons can discover their risk to employ mitigation strategies or lifestyle changes to delay dementia onset.

The proposed method has the ability to predict and model risk of cognitive health issues through personalized risk assessments to promote cognitive health for middle-aged and elderly populations.

\section{Methods}

\subsection{Data Compilation}

The proposed method utilized three categories of data in order to produce dementia risk through the machine learning algorithm: genetic data, environmental data, and lifestyle factor data subsequently trained upon cognitive disorder hospitalization data. 

Risk from environmental factors was separated into three indicators: air pollution, water pollution, and noise pollution. These types of pollution were selected as they have existing ties to cognitive disorder [27]. 
Air pollution has been widely regarded as a risk factor for developing cognitive impairment, including ADRD and Parkinson’s Disease [11]. Although the exact mechanisms for action are unknown, it is suggested that pollutants, especially that of minute particulate matter, may play a role in causing inflammation and oxidative stress, which ultimately affects the nervous system [24, 47]. The specific type of air pollution included was that of PM 2.5, or particulate matter with a width of 2.5 microns or smaller [57]. Emissions from the combustion of gasoline, wood, and oil, among others, are primarily responsible for the rising amount of PM 2.5 in the atmosphere. Such particulate matter was chosen because it is especially dangerous as it can enter the body easily [47, 57]. The dataset was provided by the Environmental Protection Agency (EPA), and each row included the State ID, County ID, and Mean PM 2.5 Concentration Level organized by county. The data itself contained more than 1,000,000 entries with Daily PM 2.5 Concentrations from 2015 to present day [18]. 

Water pollution has also been linked to cognitive impairment. Contaminants found in drinking water such as aluminum have been shown to potentially play a role in the pathogenesis of Alzheimer’s Disease [53]. However, the cellular mechanisms of aluminum in the body and its contribution to Alzheimer’s Disease still needs to be explored further [44, 53]. The presence of heavy metals such as zinc, cadmium, chromium, manganese, and iron in sources of drinking water poses a significant danger to the wildlife inhabiting the surrounding environment, disturbing the ecosystem and compromising its sustainability [12, 44]. The detrimental effects of such pollution extend to humans as well, as they contribute to further cognitive dysfunction through increased oxidative stress and the impairment of relevant enzymes. Drinking water was selected as the source of pollution as many individuals obtain their water through public utility systems, and it is one of the primary ways that water enters the body [36]. The three pollutants selected for training were arsenic, per- and polyfluoroalkyl substances (PFAS), and nitrates. Arsenic was chosen to represent the heavy metal class of pollutants and is a commonly known carcinogen, although it affects other conditions as well [12]. PFAS is an emerging type of contaminant and are synthetic compounds composed of an alkyl chain connected to fluorine atoms. Linked to a variety of health risks, PFAS is found in a variety of appliances and materials [21]. Nitrates were chosen to represent the organic pollutant category. Although nitrates are beneficial to human health in low levels, they can be harmful in levels greater than 10 mg/L as per EPA regulation standards [27, 54]. In addition to cognitive disorder through oxidative stress, excessive nitrate levels have been linked to cancer, birth defects, and thyroid disease [54]. The water pollution dataset included pollutant concentrations in drinking water from 2015 to present day to maximize the number of data points. The data itself included State ID, County ID, Mean Concentration of Arsenic, PFAS, and Nitrates in Community Water Systems, which are sources of drinking water. With more than 2,000,000 entries, the data was provided by the Environmental Protection Agency (EPA). [55]

Noise pollution was the final environmental variable factored into the algorithm to predict risk. Short term exposure to excessive noise has been linked to a decreased ability to focus and remember, while chronic exposure has been linked to increased risk of dementia [17, especially as major highways are increasingly built near homes [42]. Noise pollution data included noise levels averaged per county from 2015 to present day and contained the State ID, County ID, and Mean Noise Level. This dataset contained 600,000+ entries and was sourced from the Department of Transportation (DOT). [36]
Genetic disorder data was compiled using the Gene-Protein Interaction Network (see section: GPIN below). 

Lifestyle factor variables were factored into the proposed algorithm as well. Numerous lifestyle factors can have varying degrees of effect on an individual's dementia risk. Multiple existing dementia risk prediction models taken from the literature review utilized diet and smoking habits, two of three lifestyle factors integrated into the proposed algorithm. Poor dietary health is a risk factor for many diseases, including cardiovascular disease. Diets with consistently high amounts of saturated fat, sugar, and salt can increase cholesterol levels and cause high blood pressure, which can lead to dementia onset. Unhealthy diet was quantitatively measured by an individual’s BMI (Body Mass Index), which can then be further categorized into underweight, normal, overweight, or obese depending on an individual’s sex and age. Although BMI is not an all-encompassing factor for cognitive decline, improving one’s BMI through physical activity and fitness has been shown to decrease risk of dementia. BMI’s relation to cognitive decline was taken from the University of Michigan Health and Retirement Study. Evidence that chronic smoking increases risk of developing dementia has been consistent, as studies have found that smoking worsens vascular problems that may include minute bleeding in the brain [49]. Smoking is also directly linked to increased cell inflammation and stress which can contribute to Alzheimer’s disease and related dementias [49, 58]. Smoking habits were represented by categorization as “Current”, “Never”, or “Former” as most cohort studies examining the link between smoking and dementia onset utilized this categorization. Smoking habits and cognitive decline data was taken from the Behavioral Risk Factor Surveillance System from the CDC. A lesser number of studies also took alcohol intake into account, the final lifestyle factor incorporated. Although moderate alcohol intake has not been shown to increase risk of developing dementia, excessive alcohol intake of more than 14 units of alcohol per week over longer periods reduces the amount of white matter in the brain, reducing message transmission and function in the brain [4]. Alcohol intake was measured by units of alcohol per week. Alcohol intake and dementia case data was taken from the Alzheimer’s Disease and Healthy Aging Data by the CDC as well. 

The final type of data inputted into the algorithm was a nationwide dataset of Cognitive Disorder Hospitalizations, organized by county. Separate dataset with deaths was filtered for cognitive reasons through filtration of the keywords “Dementia” and “Alzheimer’s Disease” only and appended to the hospitalizations dataset as well. Data covered cognitive cases from 2015 to present day to include more available data points. The data itself included state, county, cause of hospitalization (“Alzheimer disease and dementia”), and number of hospitalizations taken over the aforementioned time span. The dataset included 2,000,000+ entries and was provided by the Centers for Disease Control and Prevention (CDC). All data used in the proposed algorithm is publicly available data and is therefore anonymized. Data is solely used in the aggregate and cannot be traced back to any specific individual. 

\begin{figure}[H]
\centering
{\includegraphics[width=0.5\linewidth]{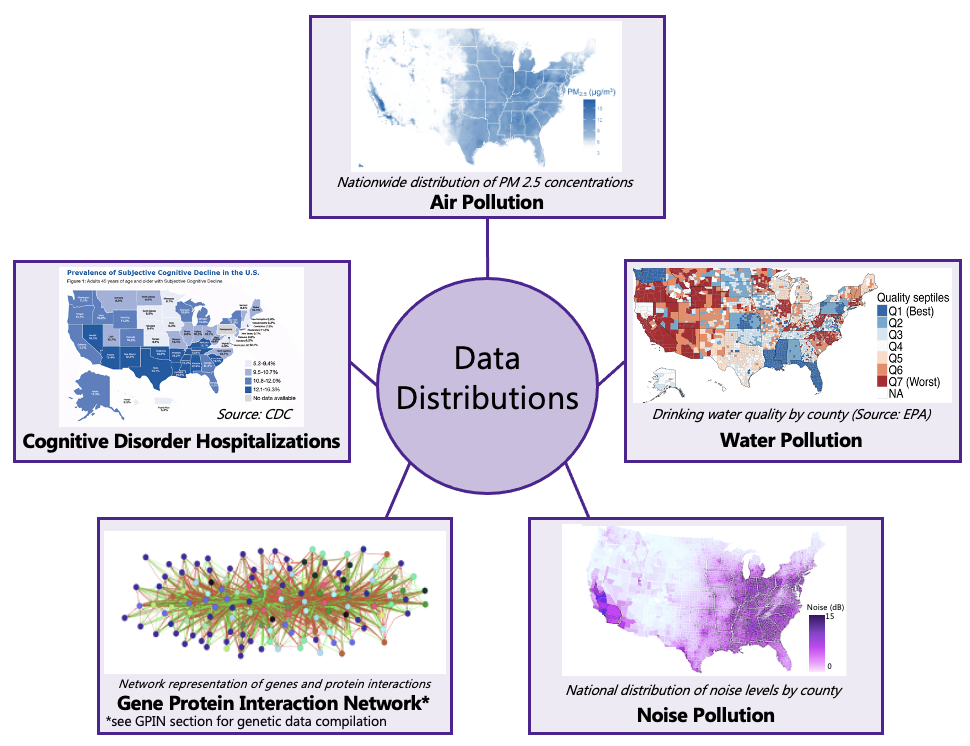}}
\caption{Nationwide distributions of various data sources within proposed algorithm}
\label{fig:falsecolor}
\end{figure}
Air pollution and noise pollution maps were created through the geologic mapping software and Python plugin Distribution Map Generator (DMG). DMG was utilized to input point localities to be plotted over an area of grid polygons of any geographical area. This research plotted on the county level as all environmental data was organized by county. Water pollution maps were sourced from the EPA and cognitive hospitalization maps were provided by the CDC. Environmental pollution distribution maps are critical to understanding the extent of the effect of environmental pollutants on cognitive decline. A comparative analysis of the featured maps concluded that higher pollution levels over the three sources generally contributed to higher cognitive decline in the southern part of the contiguous 48 states, but did not completely correlate with trends in cognitive decline in the north and east.

\subsection{Gene-Protein Interaction Network}

The Gene-Protein Interaction Network (GPIN) was created to visually represent the complex nature of gene interactions through a systems biology network as opposed to tabular data, as multimodal connections between genetic entities are often lost in the latter form of representation. 

The first step in creating the GPIN was to integrate gene and protein data to create the network. First, disease genes and genetic phenotypes were taken from the Online Mendelian Inheritance of Man (OMIM) Database. Text mining was used to extract data for neurodegenerative dementia diseases. Next, to obtain molecular targets relevant to the cognitive disorders, a pharmacological approach was taken where drugs used to treat cognitive disorders were extracted from the DrugBank database. Drug molecular targets for drugs in all developmental stages were taken, allowing for the identification of major molecular targets. Next, protein-protein interaction data was taken from the Interloguous Interaction Database (IID), and integrated dementia and drug target genes with corresponding protein interactions [34]. Finally the GPIN was created, with the final interaction network containing nodes representing disease proteins and drug targets, with edges representing protein interactions. The flowchart below shows how the GPIN was created. 
\begin{figure}[H]
\centering
{\includegraphics[width=0.5\linewidth]{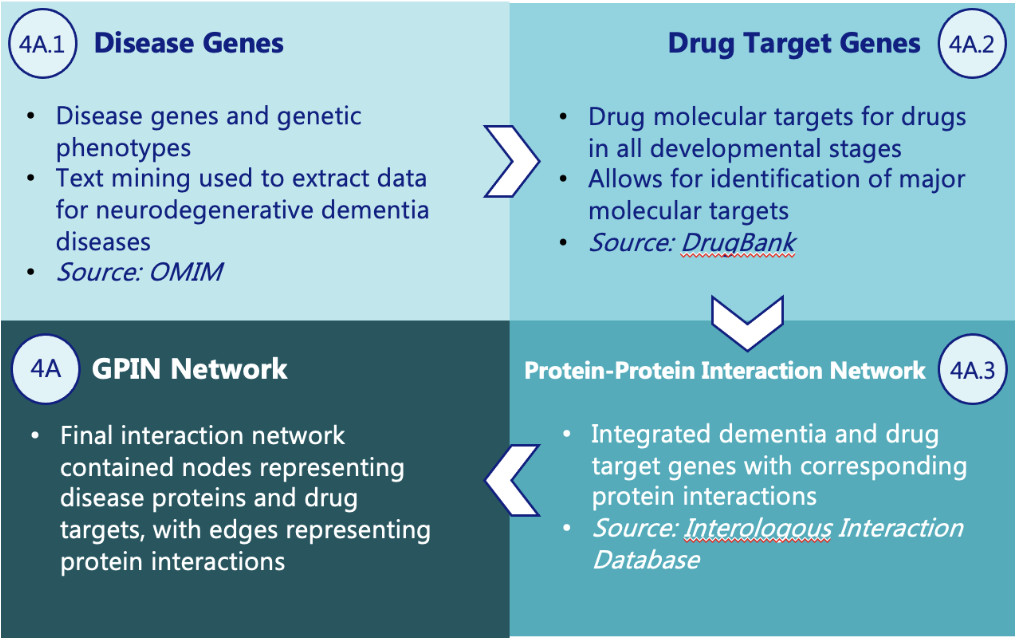}}
\caption{Creation of Gene-Protein Interaction Network}
\label{fig:falsecolor}
\end{figure}
After the creation of the GPIN, a network analysis was performed to determine which genetic entities were most linked to severe cognitive impairment, specifically ADRD. The centrality of proteins in the network was calculated to evaluate their functional importance through the degree index calculation. A graph or network $G$ can be expressed as $G(E, V)$, where $V$ is the set of vertices and $E$ is the set of edges connecting them. Edge $e_{ij}$ connects vertex $v_i$ with vertex $v_j$. In this graph, edges are undirected, so $e_{ij}$ = $e_{ji}$ . The neighborhood $N_i$ for a vertex $v_i$ is defined by its directly connected neighbors by the equation $N_i={v_j : e_j \in E}$. The degree $D_i$ of a vertex is the number of vertices satisfying the neighborhood $N_i$. Proteins were then ranked by degree index into a dataset, which is able to factor in the complexity of the interactions themselves through this network measure. Proteins with a higher degree index are more likely to be related to dementia.
Table 1 provides a condensed version of the information included in all datasets inputted into the algorithm. 

\vspace{2mm}

\begin{center}
\begin{tabular}{ |p{2.2cm}||p{3cm}|p{5cm}|p{1.6cm}|p{2cm}|  }
 \hline
 \multicolumn{5}{|c|}{Datasets Inputted} \\
 \hline
Category& Data Factor&Information Included&Number of Entries&Source \\
 \hline
 Environmental  & Air Pollution    &Mean PM 2.5 Concentration&   1,000,000+ & EPA\\
 Environmental&   Water Pollution  & Mean Arsenic, PFAS, Nitrate Concentrations   &2,000,000+ & EPA\\
 Environmental &Noise Pollution & Mean noise level (dB)&  600,000+ &  DOT\\
 Genetic &Disease Genes& Genes and Phenotypes &  16,000+ &  OMIM\\
 Genetic &Molecular Targets& Drug Gene Targets &  10,000+ &  DrugBank\\
 Genetic &Protein Data& Protein-Protein Interactions &  50,000+ &  I2D\\
 Lifestyle &Dietary Health& BMI percentage, Dementia Diagnosis, Gender &  100,000+ & University of Michigan\\
Lifestyle &Smoking Habit & Smoking Status, Race/Ethnicity, Gender, Dementia Diagnosis &  250,000+ &  CDC\\
Lifestyle &Alcohol Intake & Units of Alcohol Intake per Week, Race/Ethnicity, Gender, Dementia Diagnosis&  250,000+ &  CDC\\
Cognitive Disorder &Dementia Cases& Hospitalizations and Deaths&  1,000,000+ &  CDC\\
 \hline
\end{tabular}
\end{center}

\vspace{2mm}

\subsection{Predictor Importance Analysis}

To ensure that the correct combination of predictors was selected for training, a predictor importance analysis was performed. In line with existing dementia prediction models, the LightGBM (Gradient Boosting Machine) framework was employed to rank data inputs based upon their importance to dementia, from the combined dataset [30]. The LightGBM classifier exhibited faster gradient boosting than its XGBoost counterpart and is based upon decision tree algorithms. LightGBM performs vertical growth of the trees (depth-preferred) while XGBoost performed horizontal tree growth (breadth-preferred) [30]. 

\begin{figure}[H]
\centering
{\includegraphics[width=0.7\linewidth]{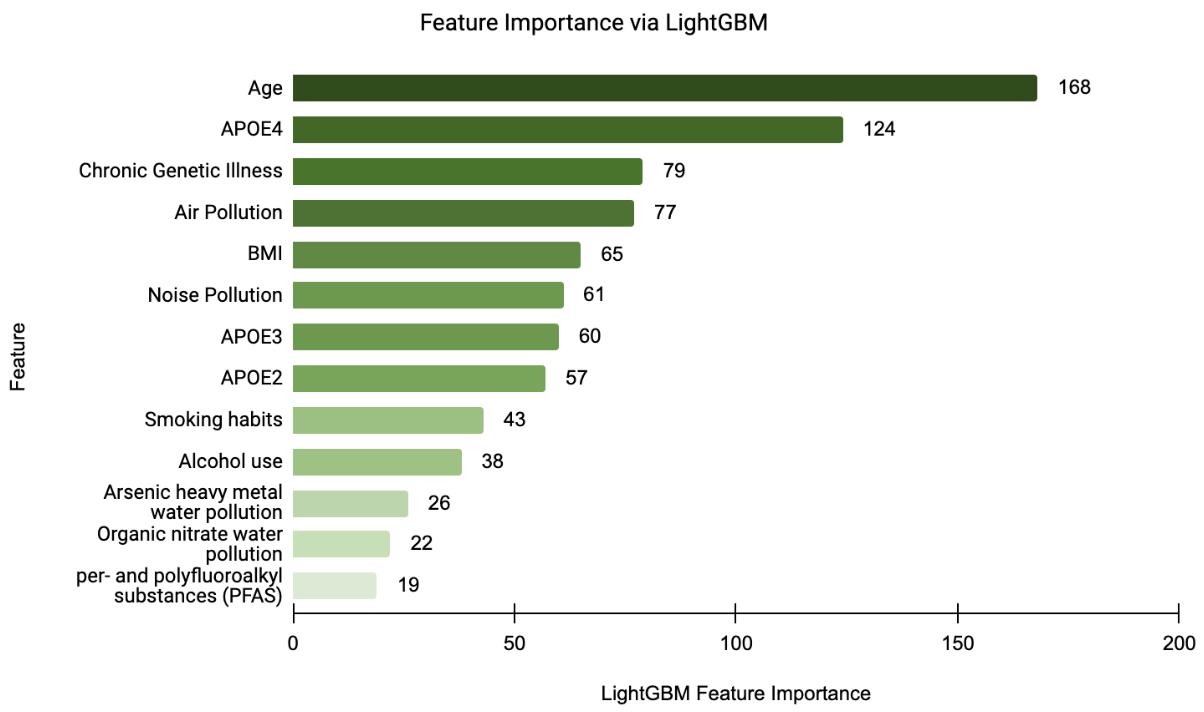}}
\caption{Predictor relevance analysis and ranking for dementia onset via LightGBM}
\label{fig:falsecolor}
\end{figure}

The above ranking of various dementia predictors aligns with existing literature on factor rankings for dementia, concluding that the chosen factors are validated for training. To account for potential multicollinearity, in which multiple variables in the statistical analysis are correlated, principal component analysis (PCA) [50] was used for dimensionality reduction and data decomposition to independent factors to reduce the effects of multicollinearity. An ANOVA test was also performed to ensure that the LightGBM feature selection was reliable to begin data integration. 

\subsection{Data Integration}

Environmental and lifestyle factor data is tabular, while the GPIN is formatted as a systems biology network. Integration of these two data formats into a robust pipeline was necessary to facilitate training. A novel tabular-network integration method, named “Sysable” was created via mathematical weighted integration methods assigning each data input a certain weight. Although weighted integration methods such as Fisher’s and Stouffer’s currently exist [26, 28], Sysable is the first to further maximize statistical power while conducting simultaneous threshold analysis on the weights themselves to prune the data during the integration itself. Not only does this simplify data organization to include only the most important factors, it reduces the size of the dataset, necessary under conditions of limited computing resource availability. 

\begin{center}

\vspace{2mm}
$S_w = \frac{3}{2*\sum_{n=1}^{k} w_i^{3}}\sum_{n=1}^{k} w_i*N*\frac{1}{\ln{P_i}}$

\vspace{2mm}

$P$ = number of values

$k$ = number of datasets

$N$ = standard normal cumulative density function

$w_i$= weight of P values in each dataset

\end{center}

\vspace{2mm}

Results of weighted integration via Sysable were then compared with degree index calculations calculated from the GPIN to ensure that genetic data had been assigned the correct weights without loss of complexity or interactions. Any discrepancies in assigned weights would then cause closeness-centrality measures on the network, in which information about the length of the shortest paths within a network was calculated by the sum of the minimal distances of a vertex to all other vertices. 

\begin{figure}[H]
\centering
{\includegraphics[width=0.6\linewidth]{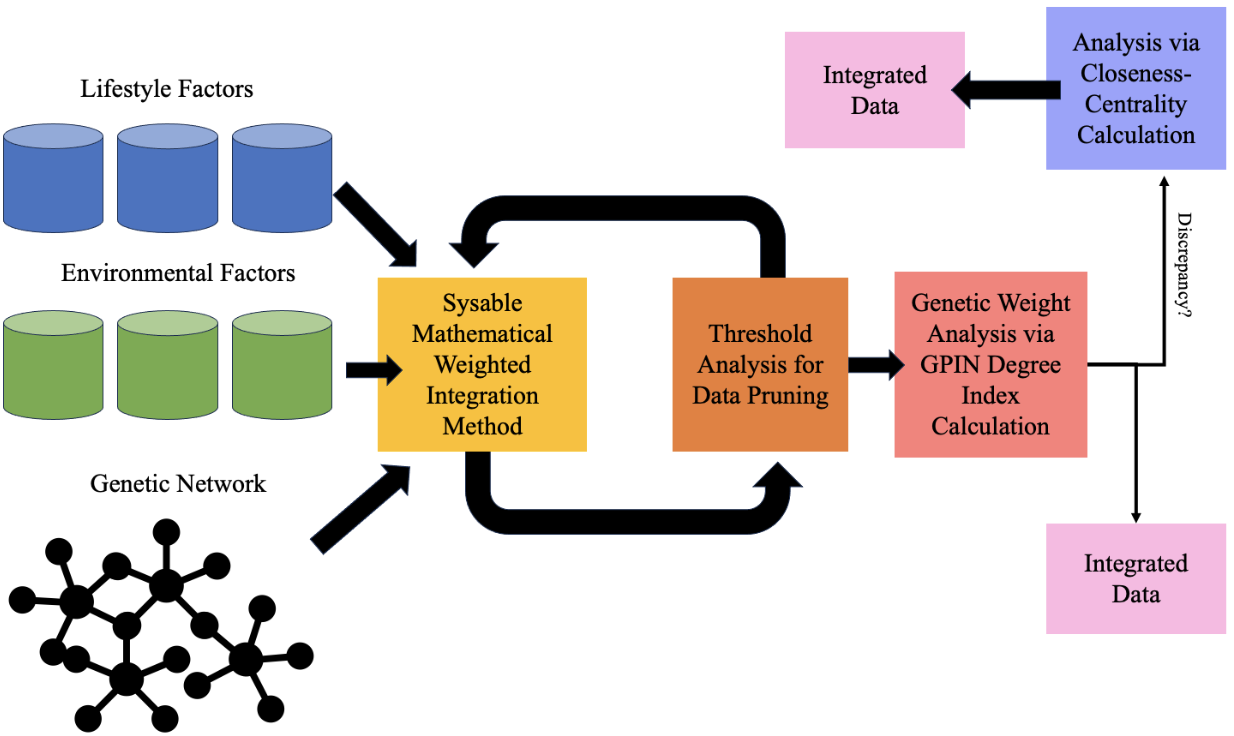}}
\caption{Flowchart diagram of tabular and network data integration via Sysable integration pipeline}
\label{fig:falsecolor}
\end{figure}

\subsection{Algorithm Construction}

The algorithm construction was framed as a “transfer learning classification problem”, allowing the algorithm to be developed with data from different populations and apply the model to another population. Transfer learning is the using of a previously learned model on a new problem, allowing neural networks to be created with relatively small amounts of data [46]. Although the datasets inputted into the model have upwards of two million data points, it dwarfs the number of elderly individuals aged 65 years and older living in the United States: 54.1 million. Furthermore, the number of middle-aged persons in the United States is more than 83 million. An approach such as transfer learning is necessary to take models and apply it to larger populations. Although transfer learning is principally used for image detection because of the substantial amount of computational power required from these tasks, it still can be applied to non-computer vision models. In addition to the ability to train models on smaller amounts of data, transfer learning has a reduced training time and reports an improved performance the majority of the time [46]. In order to implement such an algorithm to the general U.S. population through the app, transfer learning was necessary to have the results still be applicable on a wider scale. Transfer learning was performed on the proposed algorithm via the publicly available Aging Integrated Database (AGID) [15] and was evaluated with cross-validation metrics. 

\begin{figure}[H]
\centering
{\includegraphics[width=0.7\linewidth]{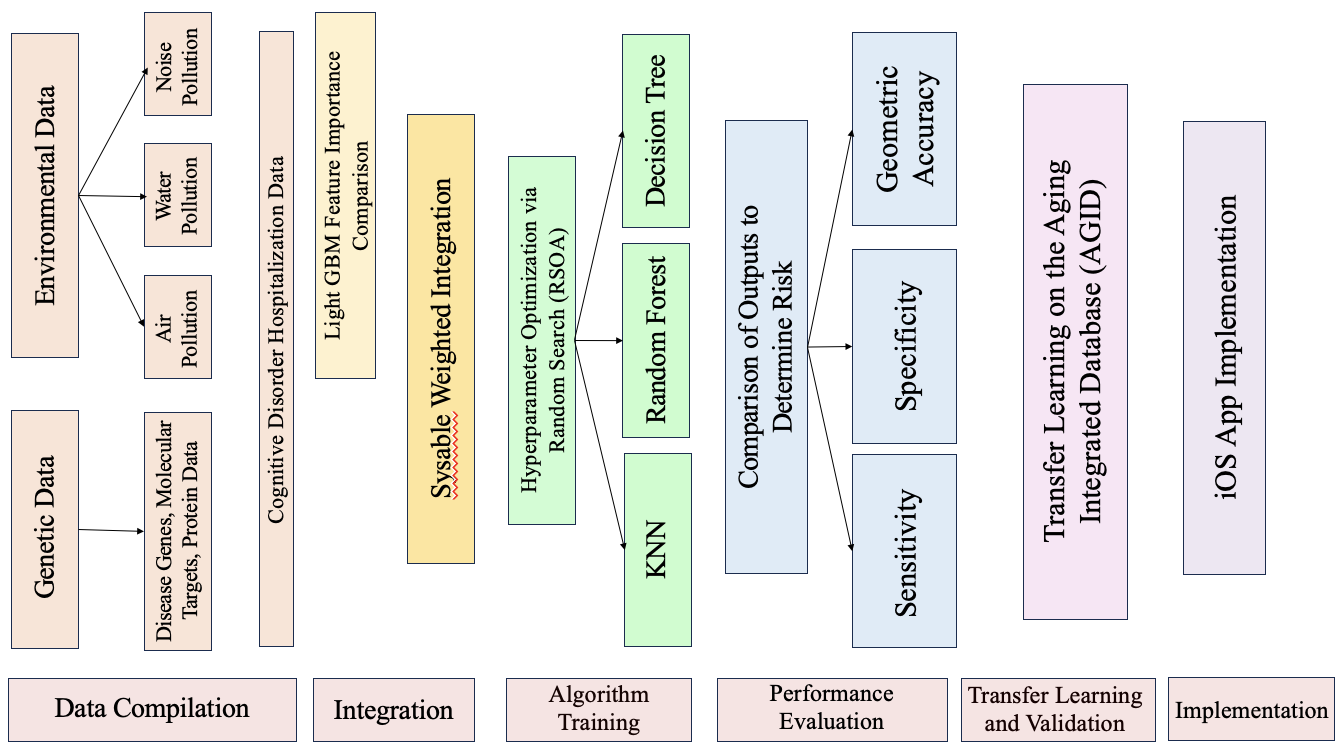}}
\caption{Proposed approach for risk computation for cognitive disorder}
\label{fig:falsecolor}
\end{figure}

The train/test split was decided to be 80\%/20\% of the complete dataset. Upon splitting, 10-fold cross validation was performed, in which the complete dataset was randomly split into 10 subsets while preserving the ratios of training and testing data. The number of folds was determined experimentally. The initial hyperparameters were then set to train 10 different models, where each model returns a different set of possible parameters. This was done to employ the Random Search Optimization Algorithm (RSOA) [7]. RSOA is used to optimize the hyperparameters that are used to train the algorithm itself. In order to train any algorithm to return accurate outputs, the functions within the algorithm need to be optimized. This optimization is often performed via one of two methods: grid search or random search. Random search generates random inputs to the objective function (the function that needs to be optimized), and is advantageous as it counteracts potential bias which is present in human-centered datasets like that of the proposed model [7]. RSOA was also selected as it is the best algorithm for noisy or non-smooth (discontinuous) areas of the search space (i.e. areas of the country with sparse or no data) [7]. RSOA is more effective than other algorithms such as grid search because it can cover a wider range of possible parameters in the same number of searches [7].

\begin{figure}[H]
\centering
{\includegraphics[width=0.4\linewidth]{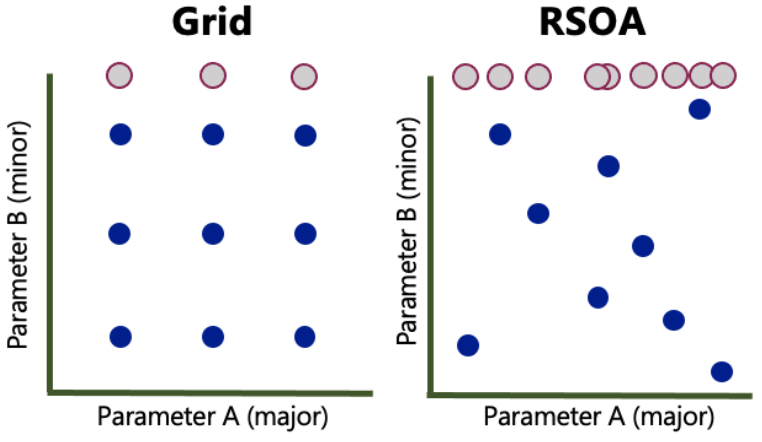}}
\caption{RSOA is more effective than other searches because it can cover a wider range of possible parameters (gray circles) in the same number of searches (blue circles)}
\label{fig:falsecolor}
\end{figure}

RSOA was further used to integrate the machine learning and systems biology approaches from the GPIN. RSOA first chose the optimal hyperparameters to train the environmental data on. However, it also can take into account the optimal set of genes to initially train the algorithm on, because the aforementioned degree index calculation and Sysable integration is able to rank genes based on their relative importance to mechanisms of action for cognitive disorders. Therefore, when the objective function is sent to the machine learning algorithms for training, genetic data is included and is therefore factored into the algorithm as well. 

\subsection{Algorithm Training}

The optimal set of hyperparameters chosen by the RSOA and including data from the GPIN was inputted into three supervised machine learning algorithms: k-nearest neighbors (KNN), decision trees, and random forest regression [8]. KNN was chosen as it is relatively straightforward and only require tuning one parameter at a time [23], the value of k, determined experimentally to be 10. K-fold cross validation was also performed. For the decision tree [40], max\_depth was to 15 and chosen experimentally. In the random forest model, 100 trees were chosen. 

The combination of different ML strategies allows for cross-checking of the dementia risk result, resulting in greater accuracy. Specifically, if risk results differed by a certain threshold, the model result that differed was thrown out and the other two risks were averaged, reducing reliance on one algorithm and increasing overall accuracy. 

\begin{figure} [H]
\centering
{\includegraphics[width=0.3\linewidth]{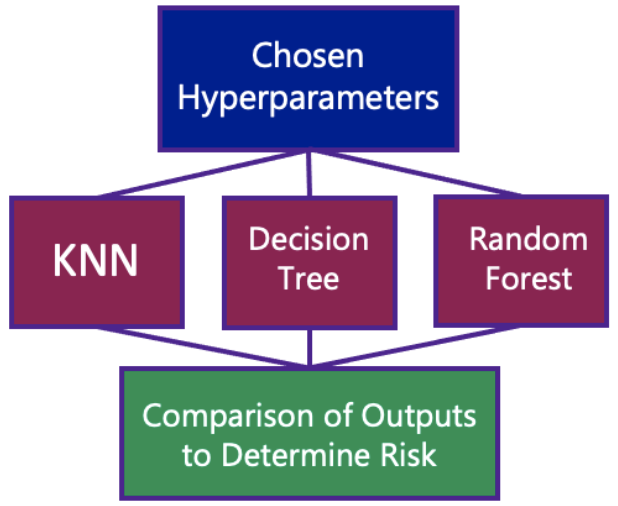}}
\caption{Flowchart of algorithm training.  }
\label{fig:falsecolor}
\end{figure}

\section{Results and Discussion}

The algorithm was evaluated on multiple metrics including sensitivity, specificity, and accuracy. These were evaluated by counting the number of true positives, true negatives, false positives, and false negatives that the algorithm outputted. Sensitivity refers to an algorithm’s ability to diagnose an individual with disease as positive (few false negatives). Specificity is the ability to identify an individual without disease as negative (few false positives). Risk was calculated by analyzing the prediction itself and the above metrics to determine the probability of a true positive (i.e. an individual truly at risk). 

\begin{figure} [H]
\centering
{\includegraphics[width=0.8\linewidth]{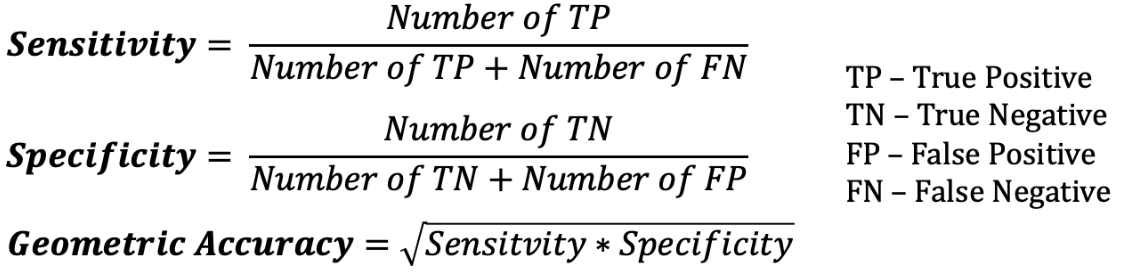}}
\caption{Formulas used to calculate sensitivity, specificity, and accuracy.}
\label{fig:falsecolor}
\end{figure}

The developed model reported a sensitivity of 85\%, specificity of 99\% and a geometric accuracy of 92\%, surpassing all other current approaches to dementia prediction. It is fundamentally different from existing approaches to dementia prediction as it is able to assimilate data from a systems biology based network and a location-based dataset. Existing approaches to cognitive disorder prediction also do not take into account the wide variety of variables that this model does, with most only focusing on genetic data for disease prediction. The developed model is the first algorithm to take both genetic and environmental data into account. 

\begin{figure} [H]
\centering
{\includegraphics[width=0.5\linewidth]{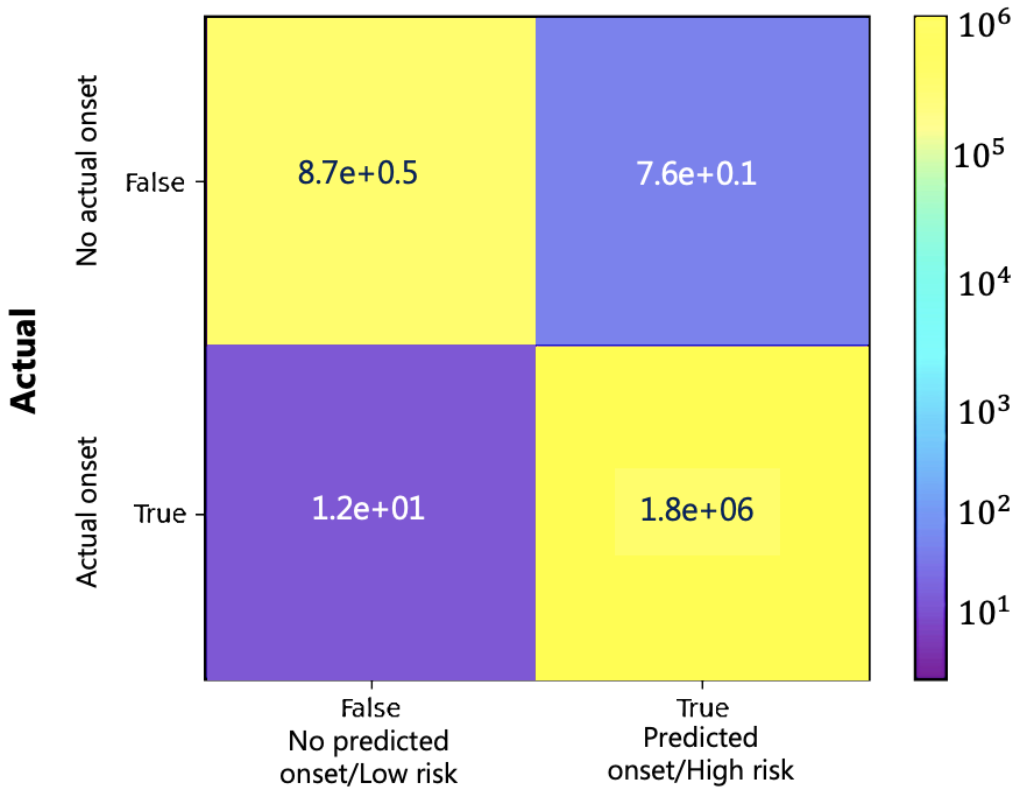}}
\caption{Confusion matrix showing prediction results compared to actual results.}
\label{fig:falsecolor}
\end{figure}

The accuracy of the developed algorithm was also modeled via the Area Under the Receiver Operating Characteristic Curve (AUROC). The ROC compares true positive rates to false positive rates and ranges from 0 to 1, and is pictured below. 

\begin{figure} [H]
\centering
{\includegraphics[width=0.4\linewidth]{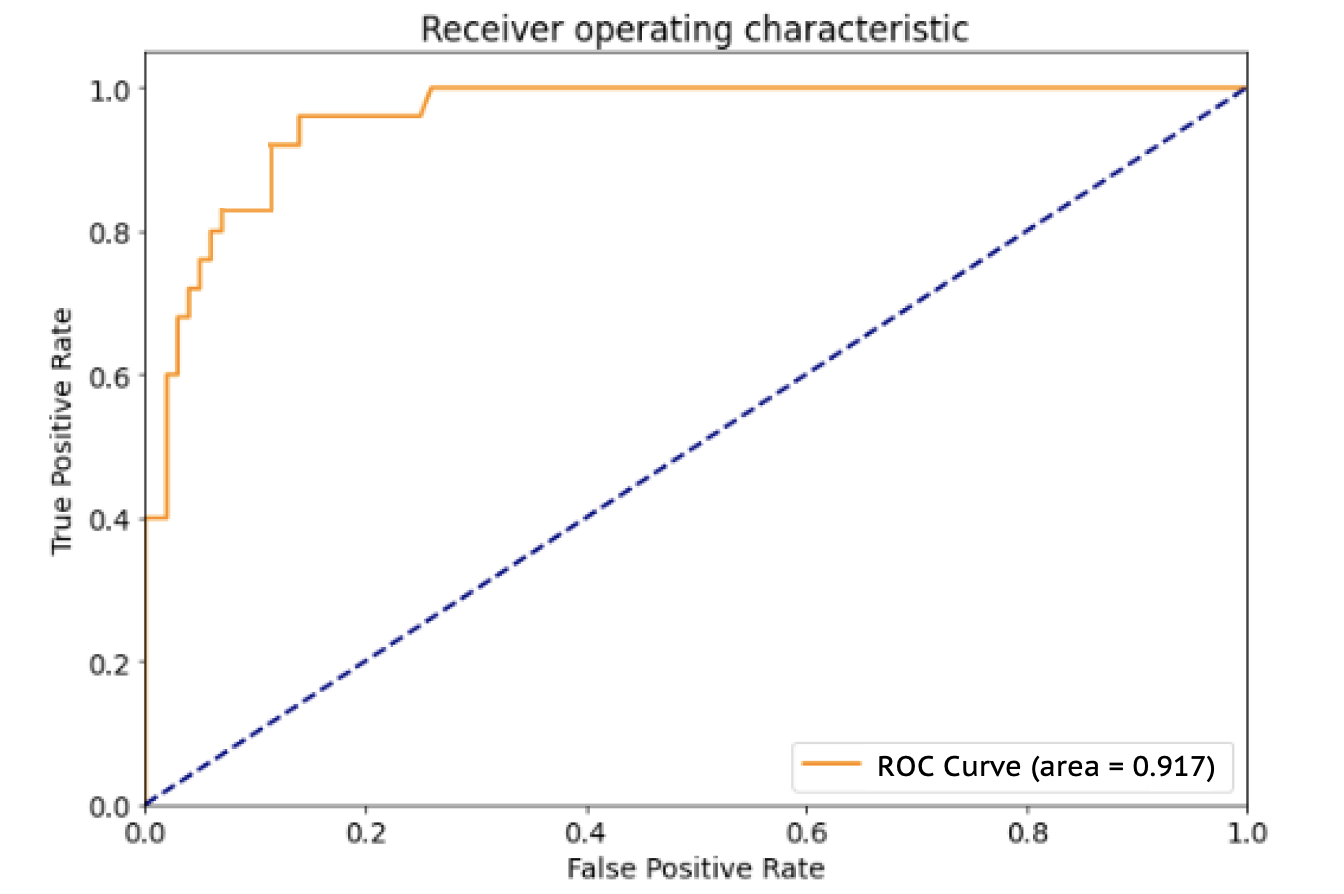}}
\caption{AUROC Curve}
\label{fig:falsecolor}
\end{figure}

Model performance of the individual machine learning algorithms on training the data were recorded to test whether the bagging, or combination, of the algorithms truly increased the accuracy of the resulting risk value. 

\vspace{2mm}

\begin{center}
\begin{tabular}{ |p{2cm}||p{2cm}|p{2cm}|p{2cm}|p{2cm}|  }
 \hline
 \multicolumn{5}{|c|}{Model-Specific Metrics} \\
 \hline
Metric& KNN&Random Forest&Decision Tree&Compound Model\\
 \hline
 Sensitivity  & 79\%    &84\%&   83\% & 85\%\\
 Specificity&   83\%  & 96\%   &93\% &99\%\\
 Geometric Accuracy &81\% & 90\%&  88\% &  92\%\\
 \hline
\end{tabular}
\end{center}

\vspace{2mm}

The performance of the model was additionally compared to existing dementia risk prediction models. The closest performing dementia risk prediction model reported by Danso et al. exhibited a geometric accuracy of 87\%, sensitivity of 76\%, and specificity of 99\%. This model was implemented on European dementia datasets and only contains genetic and lifestyle risk factors. The UKB-DRP model reported by You et al. exhibited an AUC of 0.848 ± 0.007, while the existing Cardiovascular Risk Factors, Aging, and Incidence of Dementia Risk Score reported an AUC of 0.705 ± 0.008. Additional models that the proposed algorithm definitively surpassed include the Dementia Risk Score (AUC = 0.752 ± 0.007) and Australian National University Alzheimer's Disease Risk Index (AUC = 0.584 ± 0.017). 

The proposed model underwent transfer learning from a dataset taken from the Aging Integrated Database (AGID), recently renamed as the AGing, Independence, and Disability Program Data Portal [15]. The AGID portal contains state profiles with dementia occurrence by county. State files for the contiguous 48 states were appended and trained in place of the Cognitive Disorder Hospitalizations and Deaths dataset from the CDC. The transfer model displayed a sensitivity of 79\%, specificity of 98\% and a geometric accuracy of 88\%, still surpassing all existing approaches to dementia risk prediction. 

\begin{figure} [H]
\centering
{\includegraphics[width=0.6\linewidth]{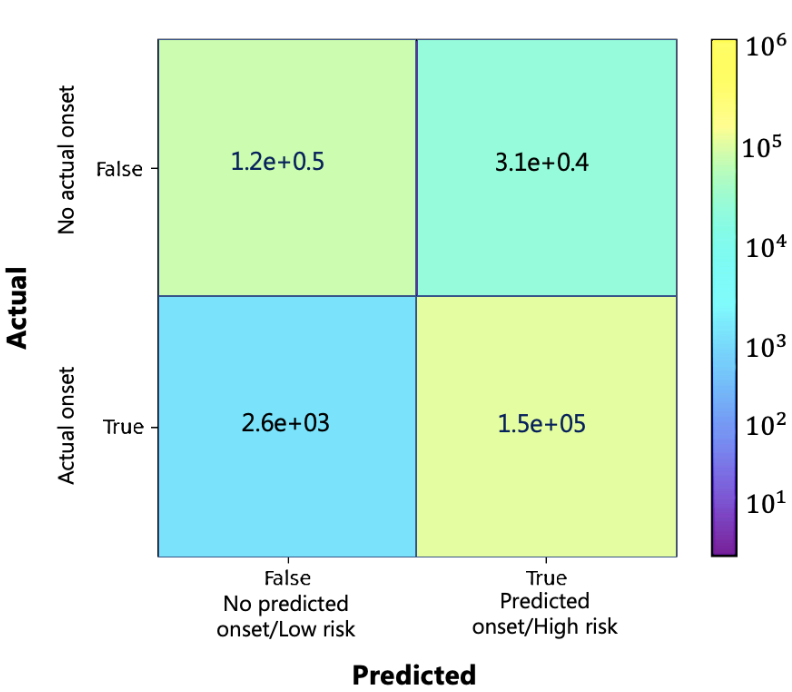}}
\caption{Confusion matrix showing transfer model prediction results compared to actual results. }
\label{fig:falsecolor}
\end{figure}

\vspace{2mm}

\begin{center}
\begin{tabular}{ |p{4cm}||p{4cm}|p{5cm}| }
 \hline
 \multicolumn{3}{|c|}{Trained vs. Transferred Model Metrics} \\
 \hline
Metric& Trained Model (CDC)&Transferred Model (AGID)\\
 \hline
 Sensitivity  & 85\%    &79\%\\
 Specificity &   99\%  & 98\% \\
 Geometric Accuracy &92\% & 88\%\\
 \hline
\end{tabular}
\end{center}

\vspace{2mm}

Additional statistical tests were performed on the trained model and transferred model, including cross-validation metrics. The table below provides statistical results from the trained model and transferred model. 

\begin{center}
\begin{tabular}{ |p{4cm}||p{4cm}|p{5cm}| }
 \hline
 \multicolumn{3}{|c|}{Trained vs. Transferred Statistics} \\
 \hline
Metric& Trained Model (CDC)&Transferred Model (AGID)\\
 \hline
 Precision  & $0.99 \pm 0.005$    & $0.82 \pm 0.008$\\
 Recall &   $0.99 \pm 0.003$  & $0.98 \pm 0.002$ \\
 F-1 Score & $0.98 \pm 0.008$ & $0.89 \pm 0.002$\\
 \hline
\end{tabular}
\end{center}

\subsection{Bias Detection}

Bias present in machine learning algorithms can pose serious problems due to objectionable discrimination. This issue is especially pertinent to the proposed model as much of the data is human centered, and any occurrence of bias can skew the results, leading to misleading results and potentially detrimental consequences for middle-aged and senior citizens. Therefore, the AI Fairness 360 Library (AIF360) from IBM was employed to detect potential bias. AIF360 is an open source toolkit and Python-based package encouraging users to “examine, report, and mitigate discrimination and bias in machine learning models throughout the AI application lifecycle” [2]. The AIF360 toolkit is applicable to this model as its fairness metrics cater to allocation or risk assessment tasks with well-defined attributes. The specific fairness metric employed the ClassificationMetric class, which applies metrics on the model itself rather than the data itself. Applying metrics on the data was unnecessary as it was publicly available and not self-collected. 

Bias detection algorithms were run in addition to the aforementioned metrics. Algorithms targeting the pre-processing, in-processing and post-processing sections of the model were run. Pre-processing bias detection algorithms are considered the most important as early intervention is usually regarded as the best intervention as it addresses group and individual fairness. Other algorithms run in the later stages included adversarial debiasing algorithms (in-processing) and two equalized odds post-processing algorithms (post-processing). 

The proposed algorithm displayed low amounts of bias and a fairly high amount of fairness, which is expected as much of the input data is taken from governmental agencies that have explicitly combed the utilized data for potential bias and unfairness. 

\subsection{Dementia Treatment Inequity Analyses}

Inequalities in cognitive disorder risk present a systemic barrier in individuals looking to seek treatment. Currently, African Americans are 1.5 to 2 times more likely than whites to develop cognitive disorders, but 35\% less likely to be diagnosed [29, 31]. Similar trends hold true for other minority races/ethnicities, as shown below in studies conducted by multiple organizations. A detailed analysis of the factors included in the dataset was performed to reveal trends in the risk predictions outputted by the proposed algorithm to help show efficient allocation of medical resources to areas most in need. 

\begin{figure} [H]
\centering
{\includegraphics[width=0.8\linewidth]{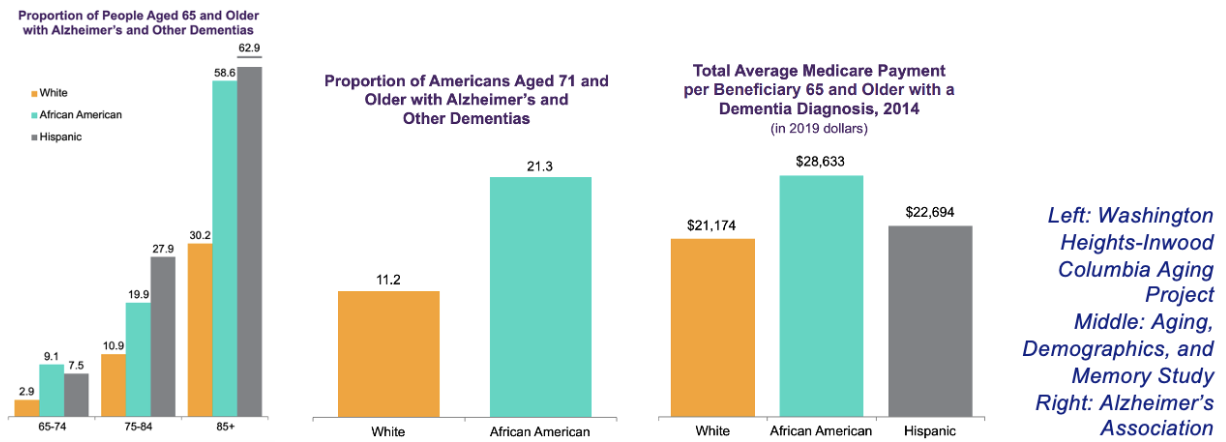}}
\caption{Graphs showing the effect of racial inequality on cognitive disorder prevalence and treatment.}
\label{fig:falsecolor}
\end{figure}

Statistical analyses were performed comparing demographics and properties of a geographical location along with the dementia prevalence taken from the input dataset. Many factors analyzed were also current healthcare disparities as reported by the CDC [5]. The selected factors were also correlated with certain environmental variables to determine if environmental conditions bear any connection to disparities in treatment. Income, education, and poverty rate data were taken from the U.S. Department of Agriculture’s Economic Research Service (USDA ERS) [9,13]. Counties with limited information were consolidated in the following scatterplots.

\begin{figure*}[ht!]
            \includegraphics[width=.33\textwidth]{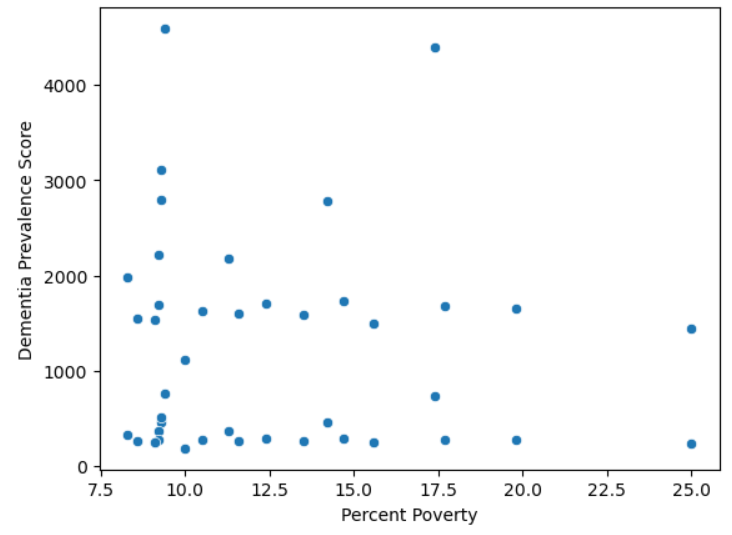}\hfill
            \includegraphics[width=.33\textwidth]{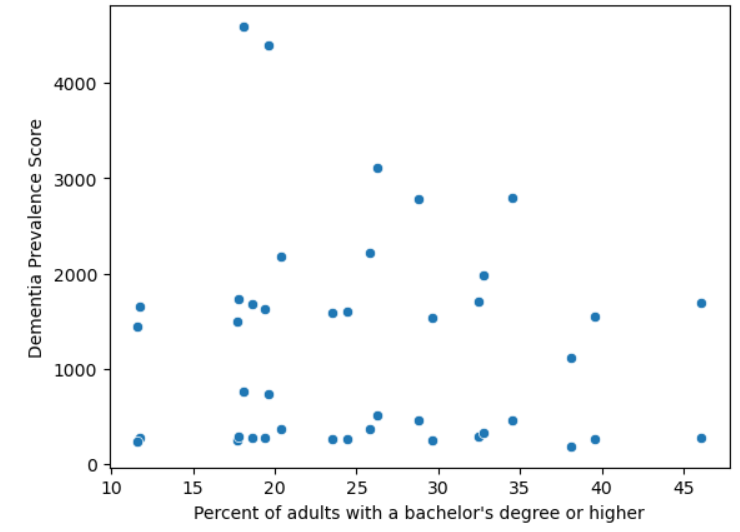}\hfill
            \includegraphics[width=.33\textwidth]{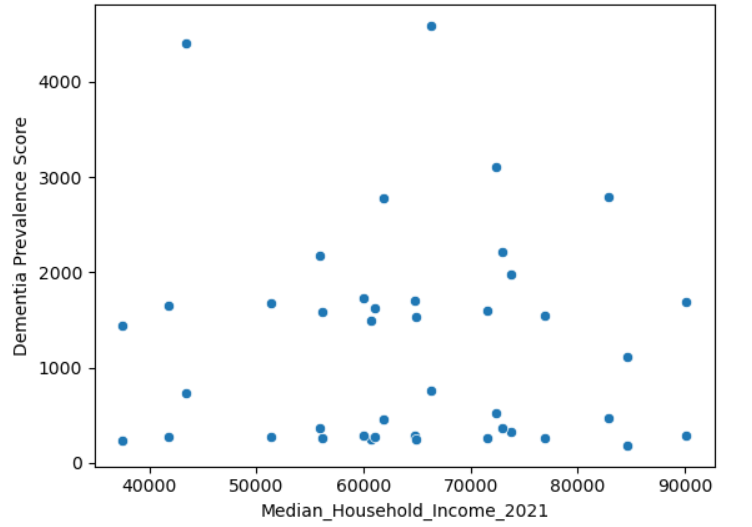}
            \caption{Consolidated County Poverty Levels vs. Dementia Prevalence}
            \caption{Consolidated County Education Levels vs. Dementia Prevalence}
            \caption{Consolidated County Income Levels vs. Dementia Prevalence}
\end{figure*}

It can be seen that there is a weak or incomplete correlation between poverty, education, or income levels and dementia prevalence. However, it should be noted that in the leftmost (poverty level) scatterplot, most points with low dementia prevalence are clustered in lower levels of poverty. In the education level scatterplot, low levels of dementia prevalence are equally distributed among education levels. The rightmost scatterplot (median income) shows nearly no correlation between income and dementia prevalence. The lack of complete correlations, however, signifies that these variables are not telling indicators of dementia onset. Analysis of one variable cannot accurately predict dementia; a holistic model taking multiple variables into account is necessary, as the proposed model does [6]. 

Potential limitations of the proposed algorithm include possible overfitting or underfitting of the data, which will cause a greater number of false negatives or false positives than the true value, respectively. Another limitation is that prediction accuracy is limited to within the US, because only environmental and cognitive hospitalization data within the US was inputted. Programs such as the Survey of Health, Ageing, and Retirement in Europe (SHARE) and PREVENT Dementia program can provide data and facilitate predictions for European populations, and is a possible next step [20,33, 35] . Next steps include adding more environmental factors such as heat and temperature data and add lifestyle factors to view the correlationary strength between these factors and severe cognitive impairment. Such trends are necessary to view in order to develop strategies to mitigate risk. 

The developed model has the potential to be utilized beyond the realm of cognitive disorders, and can assist younger populations in combining genetic and environmental factors to determine risk for chronic disease. It can also be used to combat the mental health crisis in which increasing numbers of teens are facing mental health issues such as depression and anxiety, which have both genetic and environmental influences. Using this model to determine risk from a youth’s environment is instrumental in getting them the help they need and could potentially be life-saving [1]. Beyond providing individuals with valuable information about their cognitive health, the proposed algorithm can be used to predict risk for other diseases such as Chronic Obstructive Pulmonary Disorder, which has environmental influences such as air pollution as well [11]. With relevant genetic and environmental data, the model’s innovative systems biology and machine learning based algorithm can predict risk for virtually any disease. 

\subsection{Interactive iOS Application}

The validated algorithm was then incorporated into an app intended for use by middle-aged persons, senior citizens, caregivers, and doctors. In the app, users are first welcomed through an interactive user interface, programmed via Swift and SwiftUI. The CoreML plugin was used to integrate the proposed algorithm into an accessible UI. [56]

\begin{figure} [H]
\centering
{\includegraphics[width=0.5\linewidth]{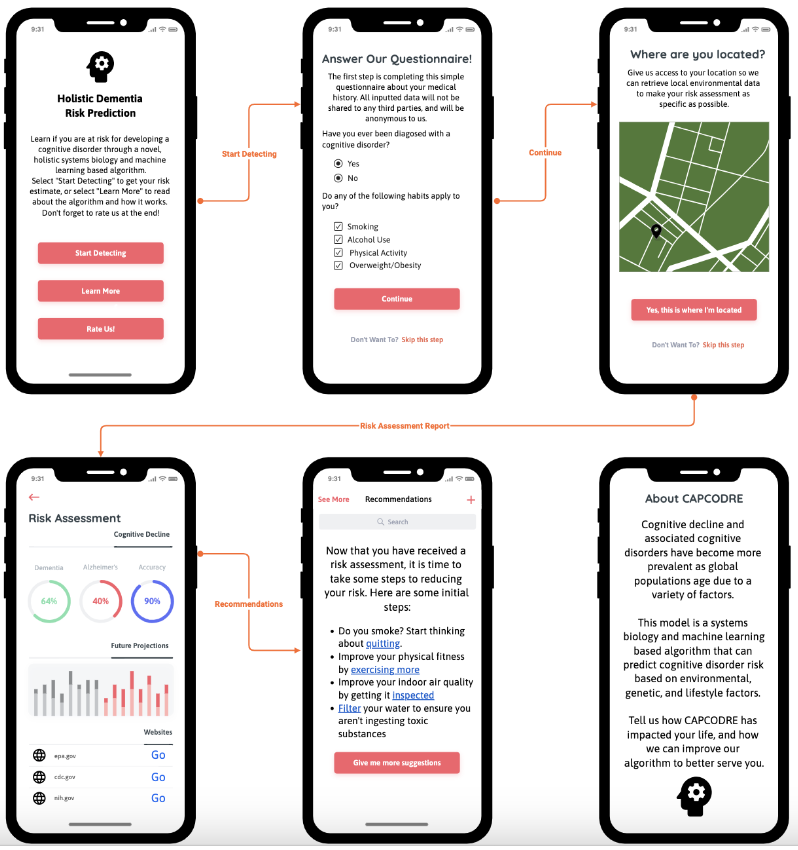}}
\caption{ Wireframe diagram of programmed app to return personalized cognitive disorder risk prediction.}
\label{fig:falsecolor}
\end{figure}

They are given the option of inputting medical history data through a questionnaire and/or allowing access to user location to gain environmental pollution data. All data is kept confidential, and data inputting is optional to maintain privacy. Cognitive disorder risk was calculated via the algorithm and presented to the user in an easily understandable and visual, graphical manner. Strategies to mitigate risk were included as well, such as tips for water or air purification, depending on the specific environmental conditions of the user. The application is particularly imperative for those in areas with limited medical care as doctors can work with patients to visualize and address risk. 

\begin{figure} [H]
\centering
{\includegraphics[width=0.6\linewidth]{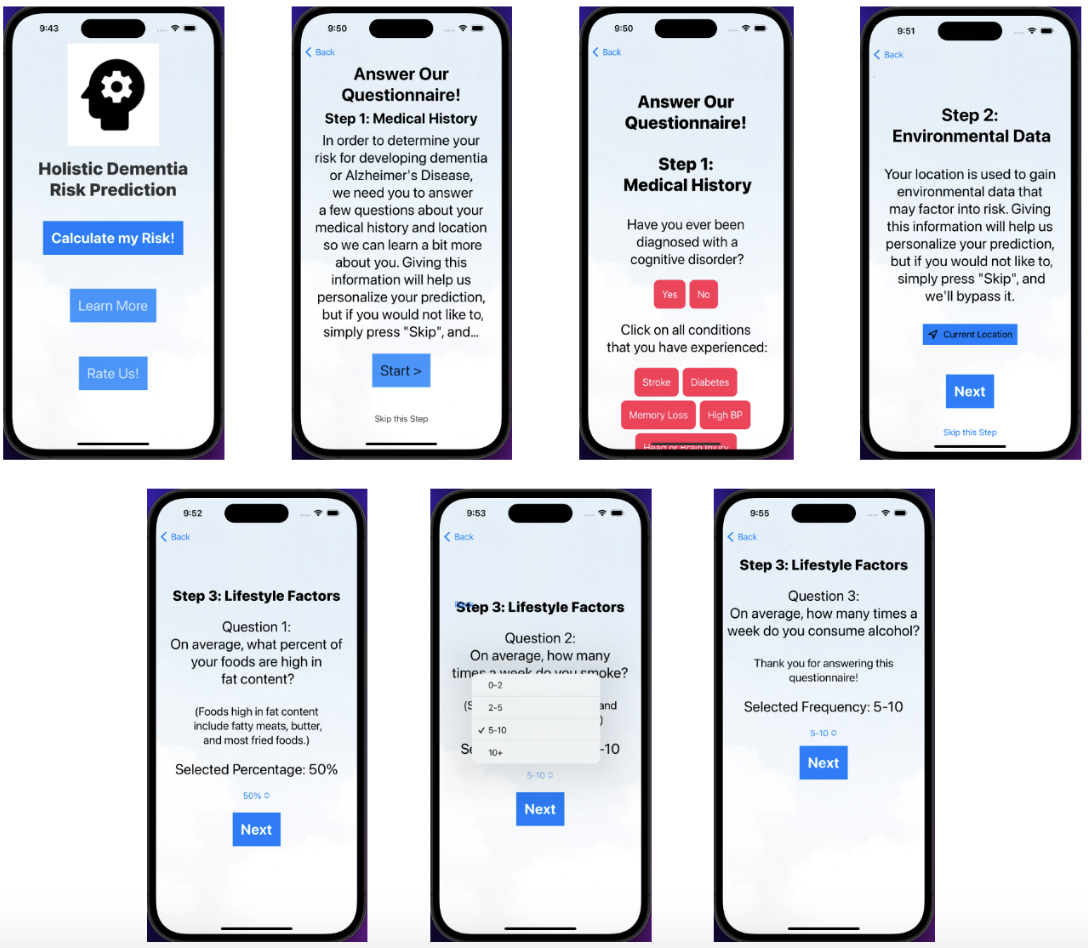}}
\caption{Screenshots of app pages guiding users through inputting environmental, genetic, and lifestyle factors. }
\label{fig:falsecolor}
\end{figure}

The programmed app is expanding to provide a pathway to treatment for senior citizens after the initial risk assessment is given. The app can serve as an assistant to doctors when differentiating between dementia and Alzheimer’s via interactive tests. One way this diagnosis is performed is through asking patients to perform various tests. The first test is often introductory in which a patient is asked common questions such as their name and the date. If a patient is able to answer correctly, doctors then ask the individual to draw a clock, and then draw in the current time. Such diagnostic tests have been shown to differentiate between dementia and Alzheimer’s as patients with the latter condition are unable to perform the task [32]. Recreating such neurological exams on a digital platform allows doctors to gain data and analytics to potentially schedule further tests through brain imaging (MRI, CT, PET).   
Mobile applications for the public to assess their risk of dementia were created by the Cardiovascular Risk Factors, Aging, and Incidence of Dementia (CAIDE) Dementia Risk Score. The CAIDE app, similar to the app involving the proposed algorithm, asks users to input various factor information and outputs a personalized risk score along with mitigation strategies to reduce risk and information for consulting doctors [45]. However, the application created from this research allows users to factor their environmental surroundings into dementia risk predictions as well. As human impact and pollution on the planet increases in prevalence and spread, environmental factors will most likely become a more significant factor in disease risk and are therefore necessary. The novel machine learning and systems biology integration of this model also allows the application to become more accurate as usage increases as the model is “learning” from each user’s input and resulting risk. Like the CAIDE application, all user input and data is strictly confidential and stored locally and cannot be accessed elsewhere. The proposed application solely predicts risk and does not make diagnoses or treatment suggestions. The application does not classify as a medical device and does not require approval from medical authorities to be used [45].

\section{Conclusion}

The proposed model is a novel computational algorithm to model and predict dementia risk in middle-aged and elderly populations. Novel holistic risk prediction was achieved through the combination of environmental pollution factors, lifestyle factors, and genetic data. Data integration of tabular and network systems biology-based data formats was performed through the novel weighted integration method Sysable. The resulting performance metrics following model training reported a higher geometric accuracy than all existing computational dementia risk prediction models. Therefore, it was concluded that the hypothesis was supported. De-biasing algorithms were run on the algorithms to ensure fairness on a human-centered dataset. Demographic disparities such as income and poverty levels were compared to dementia prevalence to visualize the effect of these factors, if any. The resulting model was incorporated into a user-friendly app for use by middle-aged and elderly populations as well as doctors. The application is especially imperative for those in areas with limited medical care. The developed framework successfully employed accurate, holistic risk prediction through computational approaches.  

\newpage

\section{References}

1. About Biomarkers and Qualification. (2021). Retrieved 3 April 2023, from https://www.fda.gov/drugs\newline /biomarker-qualification-program/about-biomarkers-and-qualification

2. AI Fairness 360, aif360.res.ibm.com/. Accessed 06 Nov. 2023.

3. Andrews, Debjani Das, Nicolas Cherbuin, Kaarin J. Anstey, Simon Easteal, Association of genetic risk factors with cognitive decline: the PATH through life project, Neurobiology of Aging, Volume 41, 2016, Pages 150-158, ISSN 0197-4580, https://doi.org/10.1016/j.neurobiolaging.2016.02.016.

4. “Alcohol and Dementia.” Alzheimer’s Society, 8 Mar. 2023, www.alzheimers.org.uk/about-dementia\newline /risk-factors-and-prevention/alcohol. Accessed 06 Nov. 2023.

5. “Barriers to Equity in Alzheimer’s and Dementia Care.” Centers for Disease Control and Prevention, 2 June 2021, www.cdc.gov/aging/publications/features/barriers-to-equity-in-alzheimers-dementia-care/index.html. Accessed 06 Nov. 2023.

6. Bennekou, Susanne Hougaard. “Moving towards a holistic approach for human health risk assessment – is the current approach fit for purpose?” EFSA Journal, vol. 17, 8 July 2019, https://doi.org/10.2903/j.efsa.2019.e170711. 

7. Bergstra, J., \& Bengio, Y. (2012). Random search for hyper-parameter optimization. Journal of machine learning research, 13(2).

8. Breiman, L. Random Forests. Machine Learning 45, 5–32 (2001).\newline  https://doi.org/10.1023/A:1010933404324

9. Bureau, US Census. “County Population by Characteristics: 2020-2022.” Census.Gov, 20 June 2023, www.census.gov/data/tables/time-series/demo/popest/2020s-counties-detail.html. Accessed 06 Nov. 2023.

10. Can genes cause dementia?. (2023). Retrieved 3 April 2023, from https://www.alzheimers.org.uk\newline /about-dementia/risk-factors-and-prevention/can-genes-cause-dementia

11. Chandra M, Rai CB, Kumari N, Sandhu VK, Chandra K, Krishna M, Kota SH, Anand KS, Oudin A. Air Pollution and Cognitive Impairment across the Life Course in Humans: A Systematic Review with Specific Focus on Income Level of Study Area. Int J Environ Res Public Health. 2022 Jan 27;19(3):1405. doi: 10.3390/ijerph19031405. PMID: 35162428; PMCID: PMC8835599.

12. Chen P, Miah MR, Aschner M. Metals and Neurodegeneration. F1000Res. 2016 Mar 17;5:F1000 Faculty Rev-366. doi: 10.12688/f1000research.7431.1. PMID: 27006759; PMCID: PMC4798150.

13. “County-Level Data Sets.” USDA ERS - County-Level Data Sets, www.ers.usda.gov/data-products\newline /county-level-data-sets/. Accessed 06 Nov. 2023.

14. Danso SO, Zeng Z, Muniz-Terrera G and Ritchie CW (2021) Developing an Explainable Machine Learning-Based Personalised Dementia Risk Prediction Model: A Transfer Learning Approach With Ensemble Learning Algorithms. Front. Big Data 4:613047. doi: 10.3389/fdata.2021.613047

15. “DATA and DATA SETS.” AlzPossible, alzpossible.org/data-and-data-sets/. Accessed 06 Nov. 2023.

16. Development of a novel dementia risk prediction model in the general population: A large, longitudinal, population-based machine-learning study, You, Jia et al., eClinicalMedicine, Volume 53, 101665

17. Do loud noises harm the brain? | Cognitive Vitality | Alzheimer's Drug Discovery Foundation. (2020). Retrieved 3 April 2023, from https://www.alzdiscovery.org/cognitive-vitality/blog/do-loud-noises-harm-the-brain

18. Download Files AirData US EPA. (2023). Retrieved 21 February 2023, from https://aqs.epa.gov/ \newline aqsweb/airdata/downloads

19. Ebaid D, Crewther SG. Time for a Systems Biological Approach to Cognitive Aging?-A Critical Review. Front Aging Neurosci. 2020 May 12;12:114. doi: 10.3389/fnagi.2020.00114. PMID: 32477097; PMCID: PMC7236912. 

20. Enable your research. (2022). Retrieved 21 February 2023, from https://www.ukbiobank.ac.uk/\newline enable-your-research

21. 'Forever chemicals' are in takeout food containers. Should you worry? | CBC News. (2023). Retrieved 3 April 2023, from https://www.cbc.ca/news/science/pfas-compostable-food-packaging-1.6794550

22. Gert Mayer, Hiddo J.L. Heerspink, Constantin Aschauer, Andreas Heinzel, Georg Heinze, Alexander Kainz, Judith Sunzenauer, Paul Perco, Dick de Zeeuw, Peter Rossing, Michelle Pena, Rainer Oberbauer; for the SYSKID Consortium, Systems Biology–Derived Biomarkers to Predict Progression of Renal Function Decline in Type 2 Diabetes. Diabetes Care 1 March 2017; 40 (3): 391–397. https://doi.org/10.2337/dc16-2202

23. Guo, Gongde \& Wang, Hui \& Bell, David \& Bi, Yaxin. (2004). KNN Model-Based Approach in Classification.

24. Hahad O, Lelieveld J, Birklein F, Lieb K, Daiber A, Münzel T. Ambient Air Pollution Increases the Risk of Cerebrovascular and Neuropsychiatric Disorders through Induction of Inflammation and Oxidative Stress. International Journal of Molecular Sciences. 2020; 21(12):4306. https://doi.org/10.3390/ijms21124306

25. Huang, Y., Sun, X., Jiang, H. et al. A machine learning approach to brain epigenetic analysis reveals kinases associated with Alzheimer’s disease. Nat Commun 12, 4472 (2021). https://doi.org/10.1038/s41467-021-24710-8

26. Hwang, Daehee, et al. “A data integration methodology for Systems Biology.” Proceedings of the National Academy of Sciences, vol. 102, no. 48, 2005, pp. 17296–17301, https://doi.org/10.1073/pnas.0508647102. 

27. Killin LO, Starr JM, Shiue IJ, Russ TC. Environmental risk factors for dementia: a systematic review. BMC Geriatr. 2016 Oct 12;16(1):175. doi: 10.1186/s12877-016-0342-y. PMID: 27729011; PMCID: PMC5059894.

28. Lapatas V, Stefanidakis M, Jimenez RC, Via A, Schneider MV. Data integration in biological research: an overview. J Biol Res (Thessalon). 2015 Sep 2;22(1):9. doi: 10.1186/s40709-015-0032-5. PMID: 26336651; PMCID: PMC4557916.

29. Lennon JC, Aita SL, Bene VAD, Rhoads T, Resch ZJ, Eloi JM, Walker KA. Black and White individuals differ in dementia prevalence, risk factors, and symptomatic presentation. Alzheimers Dement. 2022 Aug;18(8):1461-1471. doi: 10.1002/alz.12509. Epub 2021 Dec 2. PMID: 34854531; PMCID: PMC9160212.

30. “LightGBM’s Documentation!.” Welcome to LightGBM’s Documentation! - LightGBM 4.0.0 Documentation, lightgbm.readthedocs.io/en/stable/. Accessed 06 Nov. 2023.

31. “Map of Black Population, 2021.” Rural Health Information Hub, www.ruralhealthinfo.org/charts/22. Accessed 06 Nov. 2023.

32. Medical Tests. (2023). Retrieved 3 April 2023, from \newline https://www.alz.org/alzheimers-dementia/diagnosis/medicaltestsphysical

33. Modelled background pollution data - Defra, UK. (2023). Retrieved 21 February 2023, from https://uk-air.defra.gov.uk/data/pcm-data

34. Mollon, J., Knowles, E.E.M., Mathias, S.R. et al. Genetic influence on cognitive development between childhood and adulthood. Mol Psychiatry 26, 656–665 (2021). https://doi.org/10.1038/s41380-018-0277-0

35. Muirhead, M., 2015, Future potential traffic scenarios, CEDR Transnational Research Programme: Call 2012 No Deliverable 4.2, Developing Innovative Solutions for Traffic Noise Control in Europe (DISTANCE) \newline https://www.cedr.eu/download/otherpublicfiles/researchprogramme/call2012/roadnoise/distance/\newline DISTANCE-TRL-D42-V01-15072015-Future-potential-traffic-scenarios.pdf) accessed 7 July 2019

36. National Environmental Public Health Tracking Network Data Explorer. (2023). Retrieved 21 February 2023, from https://ephtracking.cdc.gov/DataExplore

37. National Institute of Health (2021) What is alzheimer's disease?, National Institute on Aging. U.S. Department of Health and Human Services. Available at: https://www.nia.nih.gov/health/what-alzheimers-disease (Accessed: April 2, 2023). 

38. National Institute of Health (2022) What is dementia? symptoms, types, and diagnosis, National Institute on Aging. U.S. Department of Health and Human Services. \newline Available at: https://www.nia.nih.gov/health/what-is-dementia (Accessed: April 2, 2023). 

39. NIH, N.I.of A. (2022) What is dementia? symptoms, types, and diagnosis, National Institute on Aging. U.S. Department of Health and Human Services. Available at: https://www.nia.nih.gov/health/what-is-dementia (Accessed: February 21, 2023). 

40. Quinlan, J.R. Induction of decision trees. Mach Learn 1, 81–106 (1986). \newline https://doi.org/10.1007/BF00116251

41. “Risk Factors for Dementia.” Alzheimer Society of Canada, alzheimer.ca/en/about-dementia/how-can-i-reduce-risk-dementia/risk-factors-dementia. Accessed 06 Nov. 2023.

42. Robbins RN, Scott T, Joska JA, Gouse H. Impact of urbanization on cognitive disorders. Curr Opin Psychiatry. 2019 May;32(3):210-217. doi: 10.1097/YCO.0000000000000490. PMID: 30695001; PMCID: PMC6438716.

43. Rowe TW, Katzourou IK, Stevenson-Hoare JO, Bracher-Smith MR, Ivanov DK, Escott-Price V. Machine learning for the life-time risk prediction of Alzheimer's disease: a systematic review. Brain Commun. 2021 Oct 21;3(4):fcab246. doi: 10.1093/braincomms/fcab246. PMID: 34805994; PMCID: PMC8598986.

44. S Shara et al 2021 IOP Conf. Ser.: Earth Environ. Sci. 623 012061

45. Sindi, S., Calov, E., Fokkens, J., Ngandu, T., Soininen, H., Tuomilehto, J., \& Kivipelto, M. (2015). The CAIDE dementia risk score app: The development of an evidence- based mobile application to predict the risk of dementia. Alzheimer's \& Dementia: Diagnosis, Assessment \& Disease Monitoring, 1(3), 328–333. https://doi.org/10.1016/j.dadm.2015.06.005

46. Sharma, P. (2021). Understanding Transfer Learning for Deep Learning. Retrieved 3 April 2023, from https://www.analyticsvidhya.com/blog/2021/10/understanding-transfer-learning-for-deep-learning/

47. Shi, L., Steenland, K., Li, H. et al. A national cohort study (2000–2018) of long-term air pollution exposure and incident dementia in older adults in the United States. Nat Commun 12, 6754 (2021). https://doi.org/10.1038/s41467-021-27049-2

48. Smith, S. (2016). 10 common elderly health issues - Vital Record. Retrieved 21 February 2023, from https://vitalrecord.tamhsc.edu/10-common-elderly-health-issues/

49. “Smoking and Dementia.” Alzheimer’s Society, 8 Mar. 2023, www.alzheimers.org.uk/about-dementia\newline /risk-factors-and-prevention/smoking-and-dementia. Accessed 06 Nov. 2023.

50. Step-By-Step Guide to Principal Component Analysis With Example, Turing Enterprises Inc, 20 July 2022, www.turing.com/kb/guide-to-principal-component-analysis. Accessed 06 Nov. 2023.

51. Tang, Eugene Y., et al. “Current developments in dementia risk prediction modelling: An updated systematic review.” PLOS ONE, vol. 10, no. 9, 2015, https://doi.org/10.1371/journal.pone.0136181. 

52. The Risk and Costs of Severe Cognitive Impairment at Older Ages: Key Findings from our Literature Review and Projection Analyses Research Brief. (2023). Retrieved 2 April 2023, from https://aspe.hhs.gov\newline /reports/risk-costs-severe-cognitive-impairment-older-ages-key-findings-our-literature-review-projection-0 

53. Trond Peder Flaten, Aluminium as a risk factor in Alzheimer’s disease, with emphasis on drinking water, Brain Research Bulletin, Volume 55, Issue 2, 2001, Pages 187-196, ISSN 0361-9230, \newline https://doi.org/10.1016/S0361-9230(01)00459-2.

54. Ward MH, Jones RR, Brender JD, de Kok TM, Weyer PJ, Nolan BT, Villanueva CM, van Breda SG. Drinking Water Nitrate and Human Health: An Updated Review. Int J Environ Res Public Health. 2018 Jul 23;15(7):1557. doi: 10.3390/ijerph15071557. PMID: 30041450; PMCID: PMC6068531.

55. Water Quality Data | US EPA. (2015). Retrieved 21 February 2023, from \newline https://www.epa.gov/waterdata/water-quality-data

56. “What Is Core ML Tools?” What Is Core ML Tools? - Guide to Core ML Tools, \newline apple.github.io/coremltools/docs-guides/source/overview-coremltools.html. Accessed 04 June 2023.

57. Wijngaarden, David Q. Rich, Wangjian Zhang, Sally W. Thurston, Shao Lin, Daniel P. Croft, Stefania Squizzato, Mauro Masiol, Philip K. Hopke, Neurodegenerative hospital admissions and long-term exposure to ambient fine particle air pollution, Annals of Epidemiology, Volume 54, 2021, Pages 79-86.e4, ISSN 1047-2797, https://doi.org/10.1016/j.annepidem.2020.09.012.

58. Zhang Q, Zhang M, Chen Y, Zhu S, Zhou W, Zhang L, Dong G, Cao Y. Smoking Status and Cognitive Function in a National Sample of Older Adults. Front Psychiatry. 2022 Jul 6;13:926708. doi: 10.3389/fpsyt.2022.926708. PMID: 35873239; PMCID: PMC9301276.

\end{spacing}
\end{document}